\documentstyle[12pt]{article}

\def\ii{\'{\char'20}}

\begin{document}
\title{Simple criterion for the occurrence of Bose-Einstein condensation}

\author{
Klaus Kirsten    \cite{kk}\\
Departament d'ECM, Facultat de F{\ii}sica
\\ Universitat de Barcelona, Av. Diagonal 647, 08028 Barcelona \\
Catalonia, Spain\\
\\and\\
\\
David J. Toms   \cite{djt}          \\
Department of Physics, University of Newcastle Upon Tyne,\\
Newcastle Upon Tyne, United Kingdom NE1 7RU}

\date{August 1995}
%\pacs

\maketitle

\begin{abstract}
We examine the occurrence of Bose-Einstein condensation in both nonrelativistic
and relativistic systems with no self-interactions in a general setting. A
 simple condition
for the occurrence of Bose-Einstein condensation can be given if we adopt
 generalized
$\zeta$-functions to define the quantum theory. We show that the crucial
feature
governing Bose-Einstein condensation is the dimension $q$ associated with the
 continuous
part of the eigenvalue spectrum of the Hamiltonian for nonrelativistic systems
 or
the spatial part of the Klein-Gordon operator for relativistic systems. In
 either case
Bose-Einstein condensation can only occur if $q\ge3$.
\end{abstract}
\eject

One of the most interesting properties of a system of bosons is that under
 certain
conditions it is possible to have a phase transition at a critical value of the
 temperature
in which all of the bosons can condense into the ground state. This was first
 predicted
over 70 years ago for the ideal nonrelativistic Bose gas \cite{Bose,Einstein}.
 The
fact that this phenomenon, now called Bose-Einstein condensation (BEC), might
explain the behaviour of liquid helium at low temperatures was suggested by
 London
\cite{London}. More recently it was suggested \cite{Blatt} that BEC could occur
for excitons in certain types of non-metallic crystals, such as CuCl. There is
 now
 good evidence for this in a number of experiments \cite{exciton}.
Another possibility for observing BEC in a real situation has arisen from the
 improved
techniques for cooling and trapping atomic systems. (See
Ref.~\cite{optical} for
 a
review.) The experimental support for BEC in cooled atoms has been advancing
 steadily
over the past few years \cite{cool}.

Given the stimulus from the various experiments which are currently taking
 place,
it is of interest to pursue a number of theoretical approaches to the problem
of
BEC in model systems. In addition, the possibility of BEC for relativistic
 systems
\cite{HW,Kapusta} is certainly of interest since BEC may play a role in
 astrophysics
and early universe
cosmology \cite{cosmo}. The most natural
setting for a discussion of BEC is within the context of quantum field theory
 where
it can be interpreted as spontaneous symmetry breaking. This was done in flat
 space
originally \cite{HW,Kapusta}, and recently in more general situations
 \cite{DJTBEC}.

We will first consider a system of nonrelativistic bosons described by a
 Schr\"{o}dinger
field $\psi(t,{\bf x})$. We will allow the space to be an arbitrary
 $D$-dimensional
manifold with a possible boundary. We also allow for the presence of an
external
electromagnetic or gravitational field, but assume that $\psi$ has no
 self-interactions.
It is possible to relate all quantities of physical interest to a knowledge of
 the
effective action $\Gamma$ which can be expressed as
\begin{equation}
          \Gamma=\ln\,{\rm det}\ell\Big\lbrack\frac{\partial}{\partial
 t}-\frac{1}{2m}
                  {\bf D}^2-\mu+U({\bf x})\Big\rbrack\label{eq:1}
\end{equation}
where $\ell$ is a constant with units of length introduced by renormalization;
${\bf D}=\nabla-ie{\bf A}$ is the gauge covariant derivative, with ${\bf A}$
the
vector potential describing any background electromagnetic fields; $\mu$ is the
chemical potential accounting for the conservation of particle number (or
 equivalently
the total charge). $U({\bf x})$ is any potential which might exist. (Instead of
 using
the effective action we could use the Helmholtz free energy or grand
 thermodynamic
potential. The relationship between these objects and the effective action, and
 a
justification for using the effective action may be found in
 Ref.~\cite{DJTnonrel}.)
We have omitted the term in $\Gamma$ which involves the classical action and
which is important for the symmetry breaking interpretation of BEC but which
 plays
no role in the following.

The formal expression in (\ref{eq:1}) requires regularization. We will adopt
 $\zeta$-function
regularization \cite{zeta} since our criterion for BEC has its simplest
 expression with this method.
The part of the effective action responsible for BEC may be expressed as
 \cite{DJTnonrel}
\begin{equation}
          \Gamma=-\zeta'(0)+\zeta(0)\ln\ell\label{eq:2}
\end{equation}
where
\begin{equation}
          \zeta(s)=\frac{T^{-s}}{\Gamma(s)}\sum_N\sum_{n=1}^{\infty}
          \frac{e^{-n\beta(\sigma_N-\mu)}}{n^{1-s}}\;.\label{eq:3}
\end{equation}
Here $T$ is the temperature and $\sigma_N$ denotes the eigenvalues of the
 Hamiltonian
operator $-\frac{1}{2m} {\bf D}^2-\mu+U({\bf x})$. The $\sigma_N$ are
recognized
as the energy eigenvalues for stationary state solutions of the Schr\"{o}dinger
 equation.

We will now assume that the bosons are confined to a space for which the
 spectrum
$\sigma_N$ splits into the sum of a discrete part $\sigma^d_{\bf p}$, and a
 continuous
part which we can deal with by imposing box normalization. The box will be
taken
to have sides $L_1,\ldots,L_q$ for some $q$, with the infinite box limit taken.
This results in
\begin{equation}
          \zeta(s)=\frac{V_q}{(4\pi)^{q/2}}\frac{T^{q/2-s}}{\Gamma(s)}
                \sum_{\bf p}\sum_{n=1}^{\infty}\frac{e^{-n\beta(
             \sigma^d_{\bf p}-\mu)}}{n^{1+q/2-s}}\;.\label{eq:4}
\end{equation}
Here ${\bf p}$ is just a set of labels for the discrete part of the spectrum,
 and
$V_q=L_1\cdots L_q$ is the volume of the box.

The condition for BEC to occur is that $\mu$ must reach a critical value
$\mu_c$
set by the lowest eigenvalue in the spectrum~:
\begin{equation}
          \mu_c=\sigma_0=\sigma^d_{\bf 0}\;.\label{eq:5}
\end{equation}
The charge in the excited states is given by
\begin{equation}
          Q_1=-eT\frac{\partial}{\partial\mu}\Gamma\;.\label{eq:6}
\end{equation}
If $Q_1$ remains bounded as $\mu\rightarrow\mu_c$ then BEC occurs, because
for the total charge large enough, it is not possible to accommodate it all in
 the
excited states. If $Q_1$ is not bounded as $\mu\rightarrow\mu_c$, then any
 amount
of the total charge can reside in the excited states and BEC will not occur. We
therefore need to look at the behaviour of
 $\frac{\partial}{\partial\mu}\zeta(0)$
and $ \frac{\partial}{\partial\mu}\zeta'(0)$ as $\mu\rightarrow\mu_c$.

Because the zero mode is playing the key role in (\ref{eq:5}), it is convenient
 to
treat it separately in (\ref{eq:4}) by defining
\begin{equation}
          \zeta(s)=\zeta^{(0)}(s)+\zeta^{(1)}(s)\label{eq:7}
\end{equation}
where
\begin{equation}
          \zeta^{(0)}(s)=
          \frac{V_q}{(4\pi)^{q/2}}\frac{T^{q/2-s}}{\Gamma(s)}
                \sum_{n=1}^{\infty}\frac{e^{-n\beta(
             \sigma^d_{\bf 0}-\mu)}}{n^{1+q/2-s}} \label{eq:8}
\end{equation}
comes from the zero mode in (\ref{eq:4}) and $\zeta^{(1)}(s)$ is given by
 (\ref{eq:4})
but with the sum over the label ${\bf p}$ restricted to the non-zero modes. For
all of the non-zero modes the argument of the exponential in (\ref{eq:4}) is
 positive
even when $\mu=\mu_c$. It is then easy to see that $\zeta^{(1)}(0)$ and
 $\zeta^{(1)}{}'(0)$
along with their derivatives with respect to $\mu$ are all finite as
 $\mu\rightarrow\mu_c$.
Whether or not BEC occurs is determined by $\zeta^{(0)}(s)$ defined in
 (\ref{eq:8}).
It is easy to see from (\ref{eq:8}) that $\zeta^{(0)}(0)=0$ and
\begin{equation}
          \zeta^{(0)}{}'(0)=
          \frac{V_q}{(4\pi)^{q/2}}T^{q/2}
                \sum_{n=1}^{\infty}\frac{e^{-n\beta(\mu-\mu_c)}}
             {n^{1+q/2}}\;.\label{eq:9}
\end{equation}
By differentiating (\ref{eq:9}) with respect to $\mu$, it is observed that
$\frac{\partial}{\partial\mu}\zeta^{(0)}{}'(0)$ diverges as
$\mu\rightarrow\mu_c$
 for
$q\le2$. There is no BEC in this case. In order that BEC occur we require
 $q\ge3$.

The restriction $q\ge3$ includes a large number of previously known results,
 often
established by long and detailed calculations, as special cases. For example,
it
 is
well known that BEC does not occur in a flat two-dimensional space
\cite{May59},
but does occur in three or more dimensions. It is known that BEC does not occur
in a finite size box in three dimensions~\cite{Pathria}. Results for BEC in a
 constant
magnetic field also follow from our result. For a constant magnetic field in
$D$
dimensions it is easy to see from the results of Ref.~\cite{DJTnonrel,DJTPLB}
that we can
identify $q=D-2p$ where $p$ is the number of non-zero components of the
magnetic
field. (This was called the effective dimension in Ref.~\cite{DJTPLB}.) Our
 result
shows that for BEC we need $D\ge3+2p$ in agreement with
\cite{DJTnonrel,DJTPLB}. In
 particular
with only a single component field we find $D\ge5$ as found originally by May
\cite{May65}. The absence of BEC for a constant magnetic field in three
 dimensions
is due to Schafroth \cite{Schafroth}. A number of other applications will be
 discussed
elsewhere \cite{KT2}.

The approach we have described for the nonrelativistic case can be applied with
a little modification to the problem of BEC in relativistic field theory. In
 place of
the Schr\"{o}dinger field we will consider a complex scalar field which may
 interact
with background electromagnetic or gravitational fields, but is otherwise free.
 Restricting
our attention to a static spacetime with the field obeying any boundary
 conditions
in the spatial directions, the effective action can again be expressed using
 generalized
$\zeta$-functions as in (\ref{eq:2}), but with $\ell^2$ in place of $\ell$ due
 to
the fact that the relevant operator in the relativistic case has dimensions of
 $({\rm
length})^{-2}$. The generalized $\zeta$-function is
\begin{equation}
          \zeta(s)=\sum_{j=-\infty}^{\infty}\sum_N\lbrack(2\pi
 Tj+i\mu)^2+\sigma_N+m^2\rbrack^{-s}\label{eq:10}
\end{equation}
where $\sigma_N$ are the eigenvalues of $-{\bf D}^2+U({\bf x})$. The energy
 $E_N$
of the mode labelled by $N$ is given by $E_N^2=\sigma_N+m^2$.

In order to extract some information from $\zeta(s)$ it is convenient to
 separate
the sum over $j$ in (\ref{eq:10}) into three pieces,
\begin{equation}
          \zeta(s)=\tilde{\zeta}(s)+\zeta_+(s)+\zeta_-(s)\;,\label{eq:11}
\end{equation}
where
\begin{eqnarray}
          \tilde{\zeta}(s)&=&\sum_N(\sigma_N+m^2-\mu^2)^{-s}\;,\label{eq:12}\\
          \zeta_{\pm}(s)&=&
          \sum_{j=1}^{\infty}\sum_N\lbrack(2\pi Tj\pm
 i\mu)^2+\sigma_N+m^2\rbrack^{-s}\;.\label{eq:13}
\end{eqnarray}
The signal for BEC is that $\mu\rightarrow\mu_c$ where
\begin{equation}
          \mu_c^2=\sigma_0+m^2=E_0^2\;.\label{eq:14}
\end{equation}
It is clear from general $\zeta$-function theory --see Ref.~\cite{Voros} for
 example--
that $\zeta_{\pm}(s)$ along with their derivatives with respect to $\mu$ will
 remain
finite as $\mu\rightarrow\mu_c$ after analytic continuation to $s=0$ is
 performed
to define $\Gamma$ and $Q_1$. We may therefore focus on $\tilde{\zeta}(s)$.

As in the nonrelativistic case the lowest mode with eigenvalue $\sigma_0$ is
 crucial
for deciding whether or not BEC occurs. If we again assume that $\sigma_N$ is
the sum of a discrete part $\sigma^d_{\bf p}$ and a continuous part which we
deal with by box normalization, then
\begin{equation}
 \tilde{\zeta}(s)=\frac{V_q}{(4\pi)^{q/2}}\frac{\Gamma(s-q/2)}{\Gamma(s)}
              \sum_{\bf p}(\sigma^d_{\bf p}+m^2-\mu^2)^{-s}\;.\label{eq:15}
\end{equation}
Since $\sigma_0=\sigma^d_{\bf 0}$ we can write
\begin{equation}
          \tilde{\zeta}(s)=\frac{V_q}{(4\pi)^{q/2}}
          \frac{\Gamma(s-q/2)}{\Gamma(s)}\Big\lbrace(\mu_c^2-\mu^2)^{q/2-s}+
              \sum_{{\bf p}\ne{\bf 0}}(\sigma^d_{\bf p}
 +m^2-\mu^2)^{-s}\Big\rbrace\;,\label{eq:16}
\end{equation}
where the sum is only over the non-zero modes. Once more general
 $\zeta$-function
theory tells us that the sum in (\ref{eq:16}) along with its $\mu$-derivative
is
 analytic
at $s=0$ even when $\mu\rightarrow\mu_c$. We can now restrict ourselves to
the first term in (\ref{eq:16}) which we call $\tilde{\zeta}_0(s)$.

The properties of $\tilde{\zeta}_0(s)$ are different depending upon whether $q$
is even or odd because of the $\Gamma$-functions which occur. For odd $q$ we
see that $ \tilde{\zeta}_0(0)=0$, and
\begin{equation}
          \tilde{\zeta}_0'(0)=
 \frac{V_q}{(4\pi)^{q/2}}\Gamma(-q/2)(\mu_c^2-\mu^2)^{q/2}\;.\label{eq:17}
\end{equation}
Because the charge $Q_1$ involves
 $\frac{\partial}{\partial\mu}\tilde{\zeta}_0'(0)$
we see that $Q_1$ remains finite as $\mu\rightarrow\mu_c$ provided that
$q\ge3$.
For $q=1$, $Q_1$ will diverge as $\mu\rightarrow\mu_c$ like
 $(\mu_c^2-\mu^2)^{-1/2}$,
and therefore BEC will not occur. The case of even $q$ results in
\begin{eqnarray}
\tilde{\zeta}_0(0)&=&\frac{V_q}{(4\pi)^{q/2}}\frac{(-1)^{q/2}}{(q/2)!}(\mu_c^2-
 \mu^2)^{q/2}\;,\\
\tilde{\zeta}_0'(0)&=&\frac{V_q}{(4\pi)^{q/2}}\frac{(-1)^{q/2}}{(q/2)!}(\mu_c^2
 -\mu^2)^{q/2}
            \lbrack\gamma+\psi(1+q/2)-\ln(\mu_c^2-\mu^2)\rbrack\;.
\end{eqnarray}
By differentiating with respect to $\mu$ it is easy to see that $Q_1$ is only
 finite
when $\mu\rightarrow\mu_c$ for $q\ge4$. If we combine the results from the
even and odd cases, we have the condition that BEC can only occur for $q\ge 3$,
exactly as in the nonrelativistic case.

As for the nonrelativistic case discussed earlier, our general result may be
applied
in a number of different situations. The flat spacetime results of
Refs.~\cite{HW,Kapusta}
are easily recovered. For a charged boson gas in a homogeneous magnetic field
described
by a single component, it was recently shown \cite{Daicic} that BEC was only
possible
if $D\ge5$. The generalization to a multicomponent field with $p$ independent
components was also performed \cite{DJTPLB}. As in the nonrelativistic case,
our
result shows that $D\ge3+2p$ is required for BEC. It is important to emphasize
that the results of Refs.~\cite{Daicic,DJTPLB} were established by lengthy
calculations,
whereas the present analysis is comparatively simple. In addition the present
analysis
shows very clearly that the condition $q\ge3$ for BEC holds for both
nonrelativistic
and relativistic scalar fields under fairly general circumstances. This was
previously
known only for free bosons or charged bosons in a homogeneous magnetic field.

In conclusion, we have shown how a criterion for BEC can be given quite simply
for fields without self-interactions by using generalized $\zeta$-functions and
studying
the lowest eigenvalue of a differential operator. This relates to the idea of
the effective
infrared dimension and finite size effects studied earlier \cite{Hu} in a
different
context. The extension of our method to study BEC in theories with
self-interactions
\cite{Benson} is of obvious
interest. Recently we have shown how this problem can be tackled very
efficiently
using $\zeta$-function methods \cite{KirstenToms}. It is very likely that the
method
described in the present paper can be extended to interacting field theory. In
particular
it is possible to define an effective field theory describing the lowest modes
as
in Ref.~\cite{Hu}. We will report on this and discuss a number of other
applications
elsewhere.

\vspace{5mm}

\noindent{\large \bf Acknowledgments}

K.K. thanks the members of the Department ECM of the University of
Barcelona, especially Emilio Elizalde, for their warm hospitality.
Furthermore, K.K. acknowledges
financial support from the Alexander von Humboldt Foundation (Germany)
and
by CIRIT (Generalitat de Catalunya).

\end{document}